\documentstyle[12pt]{article}
\begin{document}
\newcommand{\bi}{\bibitem}
\null\vskip 1cm
{}~~~~~~~~~~~~~~~~~~~~~~~~~~~~~~~~~~~~~~~~~~~~~~~~~~~~~~~~~~~~~~~~~{\it
IFA-FT-395-1994}
\vskip 2cm
\begin{center}
{\bf {\Large{STRUCTURE DEPENDENCE IN }}}
\end{center}
\begin{center}
{\Large$\bf\pi\rightarrow\bf\ell\nu,\pi\rightarrow\ell\nu\gamma$}
\end{center}
\begin{center}
{\bf {\Large AND $\pi^0\rightarrow\gamma\gamma$  DECAYS}}
\end{center}
\vskip 2cm
\centerline{\bf{ \large {D.Ghilencea, L.Micu
\footnote{E-mail address:~~
MICUL@ROIFA.BITNET}}}}\vskip 3mm
\centerline{\large Institute of Atomic Physics}\centerline{\large Theoretical
Ph
ysics Department}

\centerline{\large Bucharest POBox MG-6, ROMANIA}
\vskip 2cm
\noindent
\def\am{{\cal T}(\pi^+\to l^+\nu_l\gamma)}

\begin{abstract}
 The amplitudes of $\pi$-meson decays are calculated in a
relativistic nonperturbative quark model which assumes that mesons are made
of a quark antiquark pair and of a scalar neutral component representing
the contribution of the nonelementary fluctuations of the quark gluonic
field. The experimental data can be fitted in a satisfactory manner
using current quark masses $m_u\approx1.5\  MeV, m_d\approx9\ MeV$.
\end{abstract}
\vfill\eject
\noindent
\section {Introduction}

\noindent
 The existence of some uncertainties both in the experimental and in the
theoretical information concerning the form factors in the amplitudes
of various pion decays [1] stimulated the development of a plethora of models
for these processes.

The most popular are the quark-loop models,
whose central problem is the definition of the quark-meson coupling.
In some of these models the quark-meson  coupling is introduced via
effective Lagrangians and PCAC [1,2,3]. These models have a hybrid
character due to the simultaneous presence of quarks and mesons.
More consistent than these models are those based on the Bethe
Salpeter equation [4], which assume that a meson is a bound state of a
quark and an antiquark; however, they have some defficiences related to
the form of the interaction kernel, the validity of the instantaneous
approximation and the appearance of the relative time variable.\par
 Recently, without defining any special binding mechanism, simple models of
mesons, like products of free quark and antiquark states with a suitable
relativistic coupling of the spins and a particular
distribution of the relative momentum have been proposed [5]. Horbatsch and
Koniuk went a step further by adding a gluon to the $q\bar q$  state [6] in
order to incorporate some confining effects and possible gluon exchange.
Some time ago we also proposed [7] a form of this kind, but, instead of
the gluon, we introduced a scalar vacuum-like field $\Phi$ intended to
represent globally the fluctuations of the quark-gluonic field which
cannot be described in terms of a few elementary excitations.\par
 Apppearing under different names like condensates in the Shifman, Vainshtein
and Zakharov approach [8], or bag in the the bag models [9], nonelementary
excitations have been formely introduced in many quark models. In deep
inelastic processes [10] they got the name of sea, because they looks like a
sea

of highly virtual particles which becomes a real source of new particles
at high energy. The main point is that it is almost always necessary to
consider besides the valence quarks some nonperturbative effects which
are essential in the description of the binding and of the generation
of new particles.\par
 In the following section we present the particular features of the model
and derive the expressions of the pion decay amplitudes. The comparison
of our results with the experimental data is performed in Section 3. In
Section 4 our conclusions are presented. The Appendix contains some
useful notations and integrals used throughout the paper, as well as
the expression of the partial width of the weak radiative decay and the
inadvertencies found by us here.\par
\vskip 2cm
\noindent
\section{  Calculation of the decay amplitudes}\par
\noindent
 The model used for calculating the decay amplitudes is based on the
following picture we have now for hadrons.\par
Hadrons look like bags, cavities or bubbles containing a few bound
valence quarks in equilibrium with the excitation of the surrounding
quark-gluonic field.\par
 The equilibrium is stable and the internal distribution of momenta does
not change as long as the hadrons remain free. During the interaction,
the equilibrium breaks down, the fluctuations of the quark-gluonic field
become very intense and new states can emerge.\par
 A long experience proves that perturbation theory is unable to describe
both the binding effects and the hadron interaction at low energy. This
fact shows clearly that the fluctuations of the quark-gluonic field
generating these effects are far from being elementary. For this
 reason we avoid considering some elementary excitations like a small
 number of  quarks
or gluons besides the valence quarks and introduce a phenomenological
field $\Phi$ to describe globally the fluctuations of the quark-gluonic
field inside the hadrons.
The field $\Phi$ has the quantum numbers of the vacuum and carries its own
momentum which is not subject of any mass shell constraint because $\Phi$
does not represent an elementary excitation.\par
  Also, in order to bypass the perturbative treatment of the interaction,
we do not start from the elementary quark gluon coupling but we prefer
looking to the time translation operators, $U(+\infty,t_0)$ and
$U(t_0,-\infty)$
,
which are more appropiate in a phenomenological, nonperturbative approach.
We describe by $U(+\infty,t_0)$ and $U(t_0,-\infty)$ the modifications
produced in the structure of the free hadronic states by an external
interaction. \par
 In the cases studied in this paper, the pion is the single hadron involved
and nothing can change its structure up to the emission of a weak or
electromagnetic quanta. Therefore, the time translation operator
$U(t_0,-\infty)$, where $t_0$ is the momentum of the first emission, acting
on the single pion state leaves it unchanged since there was no external
interaction to perturbe the internal equilibrium.\par
 In the case of the radiative decays, the emission of the second quanta
takes place in a state already perturbed by the recoil of the first emission.
Due to the smalness of the involved momenta, we assume however, that  the
perturbation of the equilibrium state can be neglected.
Therefore, the time translation operator $U(t_2,t_1)$, describing the effect
of the quantum fluctuations in the time interval between the two emissions
will also be approximated by unity.\par
According to the ideas outlined above, we propose the following expression
for the $\pi$-meson state:

$$\left\vert{M_a(P)}\right\rangle~=~{1\over(2\pi)^6}\int
d^3p~{m\over e}~d^3q~{m'\over\epsilon}~ d^4Q~\delta^{(4)}(p+q+Q-P)~f(p,q;Q)$$
$$~~~~~~~~\times\bar u({\bf p})\gamma_5 v({\bf q})\ \chi^+\lambda_a\varphi\
 a^+_\alpha ({\bf p}) b^+_\beta({\bf q})\  \Phi^+(Q) \left\vert 0 \right\rangle
\eqno(1)$$
where $a^+_\alpha(\bf p)$ and $b^+_\beta(\bf q)$ are the creation operators
of the valence quark and antiquark whose colour, spin and flavour are denoted
by the collective indices  $\alpha$ and $\beta$. The quark creation and
annihilation operators satisfy the free field  canonical
anticommutation relations, the only nonvanishing being
$$\left[ a^+_\alpha({\bf p}), a_\beta({\bf q})\right]_+=
\left[ b^+_\alpha({\bf p}), b_\beta({\bf q})\right]_+=
(2\pi)^3 {e\over m}\delta^{(3)}({\bf p} - {\bf q}) \delta_{\alpha\beta}.
\eqno(2)$$
\noindent
$\Phi^+(Q)$ denotes the creation of a nonelementary excitation having the
internal quantum numbers of the vacuum and carrying the momentum $Q$.
It represents the quantum fluctuations of the quark gluonic-field in
equilibrium with the valence quarks.It is an independent component  of the
pion wave function and hence it is supposed to commute with the quark
operators.\par
The physical meaning of the phenomenological field $ \Phi^+(Q)$ allows
us to assume that
$$\Phi(Q_1)\Phi(Q_2)\cdots\Phi^+(Q_n)=\Phi(-Q_1-Q_2+\cdots+Q_n).
\eqno(3)$$

 In the same time, keeping into account that in any process the conservation
of the quark momentum is ensured by the commutation relations (2) and that
$\Phi(Q)$ is an independent component of the hadrons, one must impose
from outside the separate conservation of its 4-momentum. We assume then
$$\left\langle 0\left\vert\Phi(Q)\right\vert 0\right\rangle=
(2\pi)^4\mu^4\delta^{(4)}(Q)
\eqno(4)$$
where $\mu$ is a constant having the dimension of mass introduced for
dimensional reasons.\par
 The binding effects in the expression (1) of the pion state are
introduced by means of the internal wave function $f(p,q;Q)$
which must cut off the large relative momenta in order to allow
the fulfilment of the orthogonality relation
$$\left\langle M_a(P)\vert M_b(P') \right\rangle =
  \delta_{ab} (2\pi)^{3}\  {E\over M}\ \delta^{(3)}(P'-P)\ \delta(M'-M).
\eqno(5)$$
 The factor $\delta(M'-M)$ in equation (5) is due to the
equation (4) following from the nonelementary
character of the phenomenological field $\Phi$.It forces us to modify
the phase space density as it follows
$${1\over(2\pi)^3}d^3 P{M\over E}\to
{1\over(2\pi)^3}dM~d^3~P~{M\over E}~\rho(M,M_0)
\eqno(6)$$
where $\rho(M,M_0)$ is the probability distribution of the pion mass
around a central value $M_0$.
If  $\rho(M,M_0)$ has a small width, like in our case,
the integral over M can be performed from the beginning and one can return
to the old phase space density.\par
We pass now to the calculation of the pion decay amplitudes.\par
\vfill\eject
\vskip 1cm\noindent
{\it a}). ${\pi^{\pm}}\to l^{\pm}\nu_l$\par\noindent
In the lowest order of the weak interaction, the $S$ matrix element of  pion
decay into leptons is
$$S(\pi^{\pm}\to l^{\pm}\nu_l)\ =~i\int
d^4x~\left\langle l^{\pm}\nu\left\vert
{}~ U(+\infty,x_0) ~H_w(x)~U(x,-\infty)\right\vert \pi^{\pm}(P)\right\rangle
\eqno(7)$$
where the weak hamiltonian
$$H_w~=~{G_F\over\sqrt 2}\cos\theta_C\left(J^{(q,w)+}_\mu(x)
J^{(l,w)-}_\nu(x)\ g^{\mu\nu}+h.c.\right)$$
$$~~~~=~{G_F\over\sqrt 2}\cos\theta_C\left(\bar d(x)
\gamma_{\mu}(1-\gamma_5)u(x)~
{}~\bar {\nu_l}(x)\gamma^{\mu}\left(1-\gamma_5\right)l(x)+h.c.\right)
\eqno(8)$$
is written in the limit of local quark-lepton interaction.In equation (8)
$G_F$ is the Fermi constant, $\theta_C$ is the Cabibbo angle, $u(x)$, $d(x)$
are the up and down quark fields and $l(x)$, $\nu(x)$ are the lepton fields.
The time translation operators $U(+\infty,x_0)$, $U(x_0,-\infty)$ in equation
$(7)$ represent the global effect of the quantum fluctuations due to some
strong external interactions which are absent both in the single pion
and in the final state where only leptons are present.Then,
as we have already mentioned above, it follows that the time translation
operators $U(+\infty,x_0)$, and $U(x_0, -\infty)$ in equation $(7)$
can be replaced by unity.\par
 After introducing in equation (7) the plane wave decomposition of the
free fields (A1,A2), the expression (7) of the $S$ matrix element
can be written as it follows
$$S(\pi^+\to l^+\nu_l)~=~-i{G_F\over\sqrt 2}\cos\theta_C~
 \overline u_{\nu_l}(l')\gamma^\mu\left(1-\gamma_5\right)v_l(l)$$
$$~~~~~~~~~\times\int d^4x~e^{i(l+l')x}~
\left\langle0\left\vert{\overline d}(x)\gamma_\mu\gamma_5 u(x)\right
\vert\pi^+(P)\right\rangle\eqno (9)$$
where $u_{\nu_l}(l')$ and $v_l(l)$ are the Dirac spinors for the leptons.\par
The matrix element of the quark current in equation (9) is calculated
using the expression (1) of the pion state.One gets easily
$$\left\langle0\left\vert\overline d(x)\gamma_\mu\gamma_5u(x)\right\vert
\pi^+(P)
\right\rangle~=~\left\langle0\left\vert~A_\mu^-(x)~\right\vert\pi^+(P)\right
\rangle$$
$$=~\int~d^3p~{m_u\over
e}~d^3q~{m_d\over\varepsilon}~d^4Q~e^{-i(p+q)x}~f(p,q;Q)
\delta^{(4)}(p+q+Q-P)$$
$$\times\left\langle0\vert\Phi^+(Q)\vert0\right\rangle~Tr\left(~{\hat p
+m_u\over 2m_u}\gamma_5{\hat q-m_d\over 2m_d}\gamma_\mu\gamma_5\ \right)$$
$$=e^{-iPx}\int d^3p{m_u\over e}d^3q{m_d\over\varepsilon}\ (2\pi)^4 \mu^4
f(p,q;
0)
\delta^{(4)}(p+q-P) $$
$$\times Tr\left({\hat p+m_u\over 2m_u}\gamma_5{\hat q-m_d\over
2m_d}\gamma_\mu\gamma_5\right)\eqno(10)$$
\noindent
where $\hat p=\gamma_\mu p^{\mu}$.\par
In equation (10), the condition (4) has been used.In this case it means that
the annihilation of the valence quark and antiquark of the pion into leptons
proceeds in a ``bare'' state, i.e. in the absence of other excitations
of the quark-gluonic field described by the phenomenological field $\Phi$.\par
 Using the definition of the pion decay constant  $F_{\pi^\pm}$, (A6),
one gets from equation (10)
$$F_{\pi^\pm}=-3i(\sqrt{2}\pi)^4\mu^4\ f_{\pi^\pm}^{(0)}\  {\sqrt{2}\pi p\over
M
}\
(m_u+m_d)\left(1-{(m_d-m_u)^2\over M^2}\right)\eqno(11)$$
where $f_{\pi^\pm}^{(0)}=f(p,q;0)$, and $p~=~
{1\over2}M\sqrt{[1-{(m_u+m_d)^2\over M^2}]~[1-{(m_u-m_d)^2\over M^2}]}$.
The factor 3 in equation (11) is due to the quantum number of colour and
is present whenever one encounters a quark loop.\par
It is worth noticing that equation (10) looks like those appearing in other
quark loop models, with the difference that the spin projectors are replaced
there  by propagators. In our model the valence quark and antiquark are real,
on mass shell particles and, because of this fact, the integral over their
momenta in equation (10) leads to the finite result (11). This is an
essential difference between the present model and other quark loop
models [2], where $F_\pi$ is infinite and hence cannot be calculated
directly in the absence of a consistent renormalization procedure.\par
Proceeding in the same manner, we calculate the neutral pion constant
$F_{\pi^0}$  defined by (A5) and obtain
$$F_{\pi^0}~=~-3i(2\pi)^4~\mu^4~f_{\pi^0}^{(0)}~{\sqrt{2}\pi\over
M}\sum_{j=u,d}
p_j m_j \eqno(12)$$
where $p_j~=~{1\over2}M \sqrt{1-(2m_j/M)^2}.$

Obviously, the values of $F_{\pi^pm}$ and $F_{\pi^0}$ coincide in the limit
of equal quark masses.

We associate the diagram in Fig. 1 to the expression of the matrix element
(10). It is important to notice that in our diagram
the quark lines correspond to projectors on the states of positive
and negative energy, not to propagators like in most quark-loop models [2].\par
\vskip 1cm\noindent
{\it b}). $\pi^\pm\to l^\pm\nu_l\gamma$\par
\noindent
In the lowest order of perturbation with respect to the weak and
electromagnetic interactions, the element of the $S$ matrix
corresponding to the decay of the positive charged pion can be written
$$S(\pi^+\rightarrow l^+\nu_l\gamma) ~=~(-i)^2~\int d^{4x}~d^4y$$
$$\left\langle l^{\pm}({\bf l})\nu_{l}({\bf l'})\gamma ({\bf k})\left
\vert~ U(+\infty,x_0)~H_w(x)~U(x_0,y_0)~H_{em}(y)~U(y_0,-\infty)~
\theta(x_0-y_0)\right.\right.$$
$$\left.\left.+~U(+\infty,y_0)~H_{em}(y)~U(y_0,x_0)~H_w(x)~U(x_0,-\infty)~\theta
(y_0-x_0)
\right\vert\pi^{\pm}(P)\right\rangle\eqno(13)$$
where the electromagnetic hamiltonian is
$$H_{em}(x)~=~e\left(J^{(q,em)}_{\lambda}(x)~+~J^{(l,em)}_{\lambda}(x)\right)
A^{\lambda}(x)$$
$$~=~e~\left[\sum_{i}~\varepsilon_i~\bar q_i(x)\gamma_\lambda q(x)-~
{}~\bar l(x)\gamma_\lambda l(x)\right]~A^\lambda(x)\eqno(14)$$
and  $\varepsilon_ie$ denote the quark charges.

Then, just like in $\pi^\pm\to l^\pm\nu_l$ case, the time translation operators
$U(+\infty,t)$ and $U(t,-\infty)$ can be replaced by unity.We also replace by
unity the operators $U(x_0,y_0)$ and $U(y_0,x_0)$ because the perturbation
produced in the equilibrium state by the emission of a quanta can be neglected,
as outlined in the Introduction, due to the smalness of the involved momenta.
We shall comment  more about this approximation in the last section.\par
The calculation of the $S$ matrix element proceeds then as usually, leading
to the following expression:
$$S(\pi^+\rightarrow l^+\nu_l\gamma) ~=~-e~{G_F\over\sqrt{2}}~\cos\theta_C~
\epsilon^\lambda~\int d^4x~d^4y~e^{ikx}$$
$$\times\left[
\left\langle l^+({\bf l})\nu_l({\bf l'})\left\vert
T\left(J^{(l,em)+}_{\lambda}(x)J^{(l,w)}_{\mu}(y)\right)
\right\vert0\right\rangle ~\left\langle 0\left\vert J^{(q,w)}_{\nu}
(y)\right\vert\pi^{+}( P)\right\rangle\right.$$
$$+\left.\
\left\langle l^{+}({\bf l})\nu_{l}({\bf l}')\left\vert J_{\mu}^{(l,w)}(y)
\right\vert 0\right\rangle\left\langle 0\left\vert
T\left(J^{(q,em)}_{\lambda}(x
)
J^{(q,w)}_{\nu}(y)\right)\right\vert\pi^{+}(P)\right\rangle\right]g^{\mu\nu}
\eqno(15)$$
where the first term is called the bremsstrahlung term and the second one
-the structure dependent term.The first one describes the emission of the
photon by the charged lepton, while the second one describes the emission
of the photon by the meson itself.\par
After using Wick's theorem to pass from time ordered to normal ordered
products of quark operators, we can write the matrix elements involving
two currents in equation (15) in the following manner
$$\left\langle l^+({\bf l})\nu_l({\bf l'})\left\vert T\left(
J^{(l,em)}_{\lambda}(x)~
J^{(l,w)}_{\mu}(y)\right)\right\vert0\right\rangle$$
$$=~-i\left\langle l^+({\bf l})\nu_{l}({\bf l'})
\left\vert~\colon\bar\nu_{l}(y)~\gamma_\mu
\left(1-\gamma_5\right)~S_F(y-x)~\gamma_{\lambda}~l(x)\colon~\right\vert 0
\right\rangle $$
$$=~-i\ \bar u(l')\gamma_{\mu}\left(1-\gamma_5\right)~
\int~{d^4k'\over(2\pi)^4}{e^{ik'(x-y)+ilx+il'y}\over\hat k'-m_l}~
\gamma_{\lambda}v(l);\eqno(16)$$
$$\left\langle0\left\vert ~T\left(J^{(q,em)}(x)_{\lambda}~J^{(q,w)}_{\nu}(y)
\right)\right\vert\pi^{+}(P)\right\rangle $$
$$=~i{2\over3}\left\langle0\left\vert\
\colon\bar{d}(y)\gamma_{\nu}\left(1-\gamma_5\right)~S_F(y-x)~\gamma_{\lambda}
u(x)\ \colon\right\vert\pi^{+}(P)\right\rangle$$
$$-i\ {1\over3}\left\langle0
\left\vert\bar{d}(x)\gamma_{\lambda}S_{F}(x-y)\gamma_{\nu}\left(1-\gamma_5
\right)u(y)\right\vert\pi^{+}( P)\ \right\rangle$$
$$=\ \int d^3p\ {m\over e}d^3 q\ {m_d\over\varepsilon}\ d^4Q\ \left\langle0
\ \left\vert\Phi^{+}(Q)\ \right\vert0\right\rangle~f(p,q;Q)\delta^{(4)}
(p+q+Q-P)$$
$$\times\left[\ {2\over3}
\int{d^4k'\over(2\pi)^4}\ e^{-ik'(y-x)-ipx-iqy}\
Tr\left({\hat{p}+m_u\over 2m_u}\gamma_{5}
{\hat{q}-m_d\over2m'}\gamma_{\nu}\left(1-\gamma_{5}\right)
{\hat{k'}+m_u\over k'^2-m^2_u}\gamma_\lambda\right)\right.$$
$$\left.\left.-\ {1\over3}\ \int{d^4k'\over(2\pi)^4}\
e^{-ik'(x-y)-ipy-iqx}\ Tr\left({\hat {q}-m_d\over 2m_d}\gamma_{\lambda}
{\hat{k'}+m_d\over k'^2-m_d^2}\gamma_{\nu}\left(1-\gamma_5\right)
{\hat{p}+m_u\over2m_u}\gamma_5\right)\right]\ .\right.
\eqno(17)$$
The bremsstrahlung and the structure dependent terms can be represented
diagramatically like in Figures (2a) and (2b).\par
After performing the integrals over the spatial coordinates and over the
internal momenta (A10,A11,A15,A20) in equations (15), (16) and (17),
we can put the the amplitude of the weak radiative decay into its standard form
[1]
$$ T(\pi^+\rightarrow l^+\nu_l\gamma) ~=~e\
{G_F\over\sqrt{2}}\cos\theta_C\ \epsilon^\lambda$$
$$\times\left\{m_l\sqrt{2}F_{\pi^+}\ \bar
u_{\nu_l}(l')\left[{P_\lambda\over(Pk)
}
-{2l_\lambda+k_\lambda+\sigma_{\alpha\lambda}k^\alpha\over2(kl)}\right]
(1+\gamma_5)v(l)\right.$$
$$+{1\over
M}\left.\left[\varepsilon_{\lambda\mu\alpha\beta}k^{\alpha}P^{\beta}\
 F_{V}+
i\left(g_{\lambda\mu}\ P\cdot k-P_{\lambda}k_{\mu}\right)\ F_A\right]
\epsilon^\lambda\ \bar u_{\nu_{l}}(l')\gamma^{\mu}
\left(1-\gamma_5\right)v(l)\right\}\eqno(18)$$
where $p={1\over2}M\sqrt{[1-(m_u-m_d)^2/M^2][1-(m_u+m_d)^2/M^2]}$,
$e=\sqrt{m_u^2+p^2}$,
$\varepsilon=\sqrt{m_d^2+p^2}$,
 $F_{\pi^+}$ is given by eq.(11) and the form factors
$F_V$ and $F_A$ express in terms of the usual
adimensional variable $x=2(Pk)/M^2$ and of the quark masses as it follows
$$F_V~=~{1\over x}f_V\eqno(19a)$$
$$f_V=2\pi\ \left[-3i(2\pi)^4{\mu}^4\ f_{\pi^+}^{(0)}\right]\left[{2\over3}
{m_u\over M}\ln{e+p\over e-p}-{1\over 3}{ m_d\over M}\ln{\varepsilon+p\over
\varepsilon-p}+{2 p(m_d-m_u)\over M^2}\right];
\eqno(19b)$$
$$F_A~=~{1\over x}f_{A}^{(1)}+{1\over x^2}f_{A}^{(2)}\eqno(20a)$$
$$f_{A}^{(1)}~=~2\pi\left[-3i(2\pi)^4 {\mu}^4f_{\pi^+}^{(0)}\right]\left[
{2p\ (m_u+m_d)\over M^2}-{2p\ (m_d-m_u)\over 3M^2}\right.$$
$$\left.-{2p\ (m_u+m_d)(m_d-m_u)^2\over M^4} -
{2m_u\over 3M}\ ln{e+p\over e-p}-{m_d\over 3M}\ln{\varepsilon+p\over
\varepsilon
 -p}
\right]\eqno(20b)$$
$$f_{A}^{(2)}~=~4\pi\left[-3i(2\pi)^4{\mu}^4f_{\pi^+}^{(0)}\right]\ (m_d-m_u)
\left[{p\over 3M^2}-{p(m_d^2-m_u^2)\over M^4}\right.$$
$$\left.-{1\over M^3}
\left({2\over3}m_u^2\ln{e+p\over e-p}-{1\over3}m_d^2\
ln{\varepsilon+p\over \varepsilon-p}\right)\right].\eqno(20c)$$
It is worthwile mentioning here that treating in a unitary manner the
bremsstrahlung and the structure dependent terms simply by using
the weak and electromagnetic hamiltonians of the standard model, one obtains
automatically the manifest gauge invariant form (18) of the
radiative decay amplitude. This result is a direct
consequence of the consistent treatment of the electromagnetic interaction.
\par
It is also important to notice that, contrary to the current opinion,
our form factors (19) and (20) depend essentially on the photon energy.
Both of them have poles at $x=0$, i.e. at vanishing  photon  energy, which
lead to an infrared divergence in the axial contribution to the decay width.
Although this fact is not surprising since it appears in the bremsstrahlung
part too, it requires a careful analysis of the soft photon limit, which is
beyond our purpose now.\par
\vskip 1.5cm
{\it c}). $\pi^{0}\to \gamma\gamma$\par
\noindent
The calculation of the neutral pion decay amplitude can be treated in the
same manner we did in the case of weak radiative decay of the charged pion.
Then, just like in that case, one has
$$S(\pi^{0}\to\gamma\gamma)~=~(-i)^2\int d^4xd^4y$$
$$\times\left\langle\gamma (k_1)\gamma (k_2)\left\vert\left[U(+
\infty,x_0)\
H_{em}(x)U(x_0,y_0)H_{em}(y)U(y_0,-\infty)\theta(x_0,-y_0)
\right.\right.\right.$$
$$\left.\left.\left.+\ U(+\infty,y_0)
H_{em}(y)U(y_0,x_0)H_{em}(x)U(x_0,-\infty)\theta(y_0-x_0)\right]
\right\vert\pi^{(0)}(P)\right\rangle$$
$$=-i\ \int e^{ik_1x+ik_2y}\varepsilon_1^{\mu}({\bf k}_1)
\varepsilon_2^{\nu}({\bf k}_2)$$
$$\times\left\langle\left\vert\left[{4\over9}\bar u(x)\gamma_{\mu}S_F(x-y)
\gamma_{\nu}
u(y)+{1\over9}\bar d(x)\gamma_{\mu}S_F(x-y)\gamma_{\nu}d(y)\right]
+\left[{x\to y}\atop {\mu\to\nu}\right]\right\vert\pi^0(P)\right\rangle$$
$$= -(2\pi)^4\delta^{(4)}(P-k_1-k_2) \sqrt 2 e^2 \sum_{j=u,d}
\sigma_j\ i
\int d^3p{m_j\over e}d^3p'{m_j\over e'}(2\pi)^4{\mu}^4\ f(p,q;0)
\varepsilon^2_j$$

$$\times\delta^{(4)}(p+q-P)\left[Tr\left(\gamma_5{\hat p+m_j\over
 2m_j}\gamma_{\mu}{\hat p-\hat k_1+m_j\over (p-k_1)^2-m_j^2}
\gamma_{\nu}{\hat p'-m_j\over 2m_j}\right)+
Tr\left(\mu\to\nu\atop k_1\to k_2\right)\right]
\eqno(21)$$
where $\sigma _u=1$ and $\sigma _d=-1$.\par
Writing the $S$ matrix element of the neutral pion decay into its usual form
$$S(\pi^0\to \gamma\gamma)~=~(2\pi)^4\delta^{(4)}(P-k_1-k_2)
T(\pi^0\to\gamma\gamma)\epsilon_{\mu\nu\alpha\beta}
\varepsilon^{\mu}(k_1)\varepsilon^{\nu}(k_2)k_1^{\alpha}k_2^{\beta}.
\eqno(22)$$
and performing the trace and the integration over the spatial coordinates
and internal momenta, one gets from (21) and (22)
$$ T(\pi^0\to\gamma\gamma)~=~
-3i(2\pi)^4{\mu}^4f_{\pi^0}^{(0)}{2\sqrt 2\pi\over M}\left[{4\over 9}
{m_u\over M}\ln{e_u+p_u\over e_u-p_u}-{1\over 9}{m_d\over M}
\ln{e_d+p_d\over e_d-p_d}\right]
\eqno(23)$$
where $e_j={1\over2} M\sqrt{1-4m_j^2/M^2}$, and $p_j=\sqrt{e^2_j-m_j^2}$.\par
\vskip 2cm\noindent
\section { Results}\par
\noindent
As it can be seen from the relations  (11), (12), (19), (20) and (23),
the expressions obtained for $F_{\pi^\pm}$,$F_{\pi^0}$,$F_A$,$F_V$ and
$T(\pi^0\to\gamma\gamma)$ depend on the quark masses and on the value
of the internal wave function in a given point $(Q_\mu=0)$.In order
to reduce as much as possible the dependence on the particular form of the
internal function, we shall consider the ratios $ T(\pi^0\to\gamma
\gamma)/F_{\pi^0}$,$f_V/F_{\pi^{\pm}}$,$f^{(1)}_A/F_{\pi^{\pm}}$, $f^{(2)}_A
/F_{\pi^{\pm}} $ which are functions of the quark masses only,
to be compared with the experimental values.\par
We recall that, contrary to most quark loop models, in our case the valence
quarks are real, on mass shell particles and their momenta must fulfill the
condition $p_\mu~+~q_\mu~=~P_\mu$, which is impossible to be satisfied by
the constituent quarks. We try therefore to fit the experimental data with
current quark masses and analyse our results in the range of masses
$0.1~MeV\ \leq\ m_u\ \leq5\ MeV$, $5 MeV\ \leq\ m_d\ \leq\ 15\ MeV$ as quoted
in Particle Data [10].\par
First we consider the function

\noindent
$\varphi_0(m_u,m_d)\ =\ \left(\ T(\pi^0\to
\gamma\gamma)\ /\ F_{\pi^0}\ \right)_{th}\ /\ \left(\ T(\pi^0\to\gamma\gamma)
\ /\ F_{\pi^0}\ \right)_{exp}$ and plot its contour lines
in this domain (Fig.3a).\par
The plot shows a very quick variation of $\varphi_0$ over the mass range.
It also shows that there is a continuous sequence of masses with $\
\rho=m_d/m_u\ \approx \ 6$ leading to a perfect agreement between our result
and the experimental one.

As concerns the form factors occuring in the  weak radiative decay, we notice
that one cannot perform a direct comparison with the experimental values
because, as pointed above, in our case the form factors $F_V$ and $F_A$
depend significantly on the photon energy, while the experimental data
[11,12] have been fitted with some constant values.\par

However, in order to test the predictive power of the model, we treat these
constant values like mean values of the form factors over the space region
where the measurements have been made.\par
Accordingly, we have to compare $\left(\bar{1\over x}\ f_V\ /\ F_{\pi^+}
\right)_{th}$ with $\left( F_V\ /\ F_{\pi^+}\right)_{exp}$ where $\bar{1\over
x}
$
is the mean value of ${1\over x}$ over the phase space region. Since there
are two different kinematical weights $SD^+$ and $SD^-$ given by the
equations (A26) and (A27) which correspond
to the left and right photons , one can define two mean values
$\left(\bar{1\over x}\right)_+$ and $\left(\bar{1\over x}\right)_-$:
$$\left(\bar{1\over x}\right)_+^2\ =\ {\int\int_\Delta\ {1\over x^2}\ SD^+(x,y)
\ dx\ dy\over\int\int_\Delta\ SD^+(x,y)\ dx\ dy};~~
\left(\bar{1\over x}\right)_-^2\ =\ {\int\int_\Delta\ {1\over x^2}\ SD^-(x,y)
\ dx\ dy\over \int\int_\Delta\ SD^-(x,y)\ dx\ dy}\eqno(25)$$
\noindent
where $y\ =\ {2(P\cdot l)\over M^2}$ [1], $SD^+(x,y)$ and $SD_-(x,y)$ are given
in the
Appendix and $\Delta$ is the volume in phase space. Following V. N. Bolotov
et al. [12], we choose $\Delta\ =\ \{ x,y\vert d\leq x\leq e;a+bx
\leq y\leq c\}$ with $a=1.;b=-0.8;c=1.;d=0.3;e=1.$ and find that
$\left(\bar{1\over x}\right)_+\ \approx\ \left(\bar{1\over x}\right)_-
\ \approx\ 1.65$.

In Fig.3b we ploted the contur lines of the function
$\varphi_V(m_u,m_d)\ =$

$\left(f_V/F_{\pi^+}\right)_{th}
/\left(F_V/F_{\pi^+}\right)_{exp}$.
Just like in $\varphi_0$ case one observes a very quick variation
over the quark mass range. Also, for $m_u\approx1.5, m_d\approx
9.$ one finds $\left(\bar{1\over x}\right)f_V\approx0.02$ which represents
a quite good agreement with the values quoted for $F_V$ [11].

We try the same procedure for the axial form factor but in this case the
comparison is not so relevant because of the existence of two contributions
to the axial form factor: one with a simple pole and the other with a
double pole at $x=0$. Defining as above $\left(\bar{1\over x^2}\right)_+$
and $\left(\bar{1\over x^2}\right)_-$ we find $\left(\bar{1\over x^2}
\right)_+\ \approx\ \left(\bar{1\over x^2}\right)_-\ \approx\ 3.15$ for
the volume $\Delta$.

As it can be seen from Figs.3c and 3d, the values of
$Mf_A^{(1)}(m_u,m_d)/F_{\pi^\pm}(m_u,m_d)$ and
$Mf_A^{(2)}(m_u,m_d)/F_{\pi^\pm}(m_u,m_d)$
are rather large for the quark masses in the above range,
where the agreement is reasonable in the previous cases. It follows then
that it is very hard to reach a small mean value for $(F_A)_{th}$ as
quoted in Particle Data [11], unless there is an almost perfect
compensation of the two contributions which are of different signs.
The compensation would roughly occur for
$\varphi_A(m_u,m_d)=\ f_A^{(2)}(m_u,m_d)/
f_A^{(1)}(m_u,m_d)\approx-0.5$, but, as it can be seen from Fig.3e, the
function

$\varphi_A(m_u,m_d)$ does not reach this value in the above range of current
quark masses. Then the difference between our prediction and the experimental
value of $F_A$  cannot be reduced under an order of magnitude, which is
quite bad. \par
\noindent
\vskip2cm
\section{  Comments and conclusions}\par
\noindent
The model presented in this paper has some attractive features when
compared with other quark models for pion decays. It is essentially
relativistic and allows the consistent treatment of the pion structure and of
the electromagnetic interaction, leading to manifest gauge invariant
amplitudes for radiative decays.

It also leads to finite expressions for the pion decay constants
$F_{\pi^\pm}$ and $F_{\pi^0}$ which could in principle be compared
with the experimental values. However, taking into account their dependence
on the internal wave function which could in principle be different
in the two cases, a comparison between $F_{\pi^\pm}$ and $F_{\pi^0}$
is not relevant for the predictive power of the model.

The dependence of the results on the particular internal function of the
pion is significantly reduced by considering the ratios $T(\pi^0\to\gamma
\gamma)/F_{\pi^0},\ F_V/F_{\pi^\pm},\ F_A/F_{\pi^\pm}$ which are functions
on the quark masses only. The first of these ratios is in agreement with
the experimental value for current quark masses in the range quoted in
Particle Data [11]. As concerns the other two ratios, a significant comparison
would require some information about the eventual dependence of the
form factors on the photon energy. In this sense, a new fit of the
measurements with $x$ dependent form factors including also the
interference term which was neglected up to now, would be most interesting
and could possibly remove the uncertainty in the quoted value of $F_A$ [12].

In the preceding section we pointed out a very quick variation of $T(\pi^0\to
\gamma\gamma)$ and of $f_V$ in the region of masses selected for the best
fit due to the presence in the expressions (19) and (20)
of logarithms of nearly zero arguments. It must be however stressed that this
 fit cannot be used for
the precise determination of current quark masses without a quantitative
analysis of the recoil effects on the internal equilibrium in
the quark system  after the first emission of a quanta.

The last comment concerns the infrared behaviour of the form factors.
As it can be seen from the Appendix, only the axial contribution to
 the structure dependent part leads to an infrared catastrophe, i.e.
to a divergence in the expression of the decay width. According
to the standard procedure used in quantum electrodynamics, the
infinities of this kind in the cross
section are eliminated by compensation with those appearing in the
interference term between the amplitude with radiative corrections and
without them [14,15]. In our case this compensation mechanism
does not work because of the particular form of the decay amplitude.
The problem deserves a further study following closely the arguments
of Bloch and Nordsieck [16] which are more appropriate in this case.
\vskip 2cm
\noindent
{\Large Appendix}

\noindent
{\it a}). Definitions and notations used throughout the paper
$$\Psi(x)\ =\ \int{d^3k\over(2\pi)^3}\ {m\over e}\ \left(a({\bf k})\
e^{-ikx}\ \bar u({\bf k})\ +\ b^+({\bf k})\ e^{ikx}\ v({\bf k})\right)
\eqno(A1)$$
$$A_\mu(x)\ =\ \sum_\lambda\int{d^3k\over(2\pi)^3}\ {1\over 2\omega_k}\
\ \left(\ e^{ikx}\epsilon_\mu^{(\lambda)*}\ a^+_\lambda
({\bf k})\ +\ e^{-ikx}\epsilon_\mu^{(\lambda)}a_\lambda({\bf k})\right)
\eqno(A2)$$
$$Tr\ \left(\gamma^\mu\gamma^\nu\gamma^\rho\gamma^\sigma\right)\ =\
4\left(g^{\mu\nu}\ g^{\rho\sigma}\ - \ g^{\mu\rho}\ g^{\nu\sigma}\ +\
g^{\mu\sigma}\ g^{\nu\rho}\right)\eqno(A3)$$
$$Tr\ \left(\gamma^\mu\gamma^\nu\gamma^\rho\gamma^\sigma\gamma_5\right)
\ =\ -4i\ \epsilon\ ^{\mu\nu\rho\sigma}\eqno(A4)$$
$$\left\langle0\left\vert\ A_\mu^a(0)\ \right\vert\pi^b(P)\right\rangle
\ =\ \delta_{ab}\ i\ F_{\pi^0}\ P_\mu\ ;\ a=1,2,3; F_{\pi^a}\ \approx\ 93
MeV\eqno(A5)$$
$$\left\langle0\left\vert A_\mu^\pm(0)\right\vert\pi^\pm(P)\right\rangle\ =\
i\ \sqrt2\ F_{\pi^\pm}\ P_\mu\eqno(A6)$$
\vskip 1cm
\noindent
{\it b}). Integrals

\noindent
The evaluation of the structure dependent terms arising from eq.(18)
implies the calculation of the following integrals:
$$\int\ d^3p\ {m\over e}\ d^3q\ {m'\over \varepsilon}\ \delta^{(4)}
(p+q-P)\ =\ I_0\eqno(A7)$$
$$\int\ d^3p\ {m\over e}\ d^3q\ {m'\over\varepsilon}\ {1\over2p\cdot k}\
\delta^{(4)}(p+q-P)\ =\ I\eqno(A7)$$
$$\int\ d^3p\ {m\over e}\ d^3q\ {m'\over\varepsilon}\ {p_\mu\over2p\cdot k}
\ \delta^{(4)}(p+q-P)\ =\ J_\mu\eqno(A8)$$
$$\int\ d^3p\ {m\over e}\ d^3q\ {m'\over \varepsilon}\
{p_\mu p_\nu\over2p\cdot k}\ \delta^{(4)}(p+q-P)\ =\ K_{\mu\nu}.\eqno(A9)$$

Elementary calculations give
$$I_0\ =\ {4\pi\ p\ m\ m'\over M}\eqno(A10)$$
$$I\ =\ {\pi\ m\ m'\over M\ \omega}\ ln{e+p\over e-p},\eqno(A11)$$
where $\omega$ is the energy of the real photon,
$e={1\over2}M(1+{m^2-m'^2\over 4M^2}$ and $p=\sqrt{e^2-m^2}$.

Using the criterion of covariance we write the integrals $J_\mu$
as it follows
$$J_\mu\ =\ A\ P_\mu\ +\ B\ k_\mu.\eqno(A12)$$
We observe that
$$P^\mu\ J_\mu\ =\ M\ e\ I\ =\ M^2\ A\ +\ M\ \omega\ B\eqno(A13)$$
$$k^\mu\ J_\mu\ =\ {1\over2}\ I_0\ =\ M\ \omega\ A.\eqno(A14)$$
\noindent
and obtain
$$J_\mu\ =\ {1\over2\omega}\ I_0\ {P_\mu\over M}\ +\ (eI-{1\over2\omega}
I_0)\ {k_\mu\over\omega}.\eqno(A15)$$
In a similar way, using covariance, one has
$$K_{\mu\nu}\ =\ a P_\mu P_\nu+b\ (P_\mu k_\nu+k_\mu P_\nu)+ck_\mu k_\nu
+d\ g_{\mu\nu}\eqno(A16)$$
\noindent
and, proceeding as above, one gets for $k^2=0$
$$P^\mu\ K_{\mu\nu}\ =\ MeI_\nu\ =\ (M^2 a+M\omega b+d)\ P_\nu+
(M^2b+M\omega c)\ k_\nu\eqno(A17)$$
$$k^\mu\ K_{\mu\nu}\ =\ {P_\nu\over2M}\ I_0\ =\ M\omega\ a\ P_\nu+(M\omega b
+d)\ k_\nu\eqno(A18)$$
$$g^{\mu\nu}\ K_{\mu\nu}\ =\ m^2I\ =\ M^2a+2M\omega b+4d\eqno(A19)$$
We solve the above system and write the integrals $K_{\mu\nu}$ as
it follows
$$K_{\mu\nu}\ =\ {e\over2\omega}I_0\ {P_\mu\over M}\ {P_\nu\over M}\ +\
\left({1\over2\omega}\ I_0\ -\ {m^2\over2}\ I\right)\ \left({P_\mu\over M}
{k_\nu\over\omega}\ +\ {k_\mu\over\omega}{P_\nu\over M}\ -\ g_{\mu\nu}
\right)\ $$
$$+\ \left[\omega^2\ \left(e^2+{m^2\over2}\right)I-e\omega\ I_0\right]
\ {k_\mu\over\omega}{k_\nu\over\omega}.\eqno(A20)$$
\noindent
{\it c}). $\pi^+\to l^+\nu\gamma$ decay width\par
\noindent
Following the classic paper of Brown and Bludman [13] and Ref.1, the amplitude
of the $\pi^+\to l^+\nu\gamma$ decay writes as follows
$$T\ =\ T_{IB}\ +\ T_{SDV}\ +\ T_{SDA}$$
where
$$T_{IB}\, =\, ie{G_F\ cos\theta_C\over\sqrt2}m_l\sqrt2\ F_{\pi^+}
\bar u_{\nu_l}\left({P_\lambda\over P\cdot k}-
{2 l_\lambda+k_\lambda+\sigma_{\mu\lambda}k^\mu\over 2k\cdot l}
\right)\epsilon^\lambda\ (1+\gamma_5)\ v_l\eqno(A21)$$
$$T_{SDV}\ =\ {eG_F\ cos\theta_C\over\sqrt2 M}\ L_\mu\ \epsilon^\nu
\ F_V\ \epsilon_{\mu\nu\rho\sigma}P^\rho k^\sigma\eqno(A22)$$
$$T_{SDA}\ =\ {ieG_F\ cos\theta_C\over\sqrt2 M}\ L^\mu\ \epsilon^\nu
\ F_A\ \left[P\cdot k\ g_{\mu\nu}\ -\ k_\mu P_\nu\right] ,\eqno(A23)$$
whith $L^\mu\ =\ \bar u^{\mu_l}({\bf l}')\gamma^\mu(1-\gamma_5)v_l({\bf l})$
and $\sigma_{\lambda\mu}={1\over2}\ [\gamma_\lambda,\gamma_\mu]$.

Taking the square of the modulus of the above amplitude and summing over
the polarisations one gets the expression of the partial decay rate
[11,13]
$${d^2\Gamma_{\pi^+\to l^+\nu\gamma}\over dx\ dy}\ =\ {d^2\Gamma_{IB}\over
dx\ dy}\ +\ {d^2\Gamma_{SD}\over dx\ dy}\ +\ {d^2\Gamma_{INT}\over dx\ dy}$$
where
$${d^2\Gamma_{IB}\over dx\ dy}\ =\ {\alpha
\gamma_{\pi\to l\nu}\over2\pi(1-r)}\ IB$$
$${d^2\Gamma_{INT}\over dx\ dy}\ =\ {\alpha\over 2\pi}\
{1\over\sqrt{2} F_{\pi^\pm}}\ \Gamma_{\pi^+\to l^+\nu}\ [(F_V-F_A)\ {\cal F}\ +
\ (F_V+F_A)\ {\cal G}]$$
$${d^2\Gamma_{SD}\over dx\ dy}\ =\ {\alpha\over16\pi}\ \Gamma_{\pi\to
l\nu}{1\over r(1-r)^2}\left({M\over F_\pi}\right)^2$$
$$\times[(F_V-F_A)^2\ SD^+\ +\ (F_V+F_A)^2\ SD^-].\eqno(A24)$$
Here
$$IB\ =\ {1-y+r\over x^2(x+y-1-r)}\ \left[x^2+2(1-x)(1-r)-{2xr(1-r)\over x+
y-1-r}\right]\eqno(A25)$$
$$SD^+\ =\ (x+y-1-r)[(x+y-1)(1-x)-r]\eqno(A26)$$
$$SD^-\ =\ (1-y+r)[(1-x)(1-y)+r]\eqno(A27)$$
$${\cal F}\ =\ {1-y+r\over x(x+y-1-r)}\left[(1-x)(1-x-y)+r\right]\eqno(A28)$$
$${\cal G}\ =\ {1-y+r\over
x(x+y-1-r)}\left[x^2-(1-x)(1-x-y)-r\right]\eqno(A29)$
$
where $r=(m_l/M)^2$, $x=2(P\cdot k)/M^2$ and $y=2(P\cdot l)/M^2$.

We point out here two inadvertencies appearing in the expression of the
partial decay width quoted in Ref.[11] and [13].\par
The first one concerns the sign of the last term in the square bracket
in $IB$, which is different from that appearing in Ref [13] and the
forthcoming papers. This difference is negligible in the case of pion
decay into light leptons, but it can be important in
other decays involving heavier leptons.
The second one concerns the interchange of $F_A$ with $-F_A$when passing
from $\pi^-$ decay  (Ref. [11]) to $\pi^+$ decay. Ref.[13] does not mention
this difference.

\vskip 1.5cm
{\Large Figure captions}

Fig. 1 Quark diagram for the pion decay constant $F_\pi$. The pion
is represented by the quark and the antiquark lines. The bubble represents the
pion internal wave function.
\vskip 1cm
Fig. 2 a) The diagram of the bremsstrahlung term;

b) Quark diagram of the structure dependent terms.
\vskip 1cm
Fig. 3 Contour plots of the functions

a) $\varphi_0(m_u,m_d)$;

b) $\varphi_V(m_u,m_d)$;

c) $Mf_A^{(1)}(m_u,m_d)/F_{\pi^\pm}(m_u,m_d)$;

d) $Mf_A^{(2)}(m_u,m_d)/F_{\pi^\pm}(m_u,m_d)$;

e) $\varphi_A(m_u,m_d)$

in the domain of current quark masses ($m=m_u;~m'=m_d$).

\end{document}